\documentclass[aps,prb,twocolumn,superscriptaddress]{revtex4-1}

\usepackage{graphicx,latexsym}
\usepackage{amsmath,amssymb,amsfonts}
\usepackage{bm}
\usepackage{braket}

\begin{document}


\title{Breakdown of topological Thouless pumping in the strongly interacting regime}
\author{Masaya Nakagawa}
\email{masaya.nakagawa@riken.jp}
\affiliation{RIKEN Center for Emergent Matter Science (CEMS), Wako, Saitama 351-0198, Japan}
\author{Tsuneya Yoshida}
\affiliation{Department of Physics, University of Tsukuba, Ibaraki 305-8571, Japan}
\author{Robert Peters}
\author{Norio Kawakami}
\affiliation{Department of Physics, Kyoto University, Kyoto 606-8502, Japan}




\date{\today}

\begin{abstract}
We elucidate the mechanism for instability of topological Thouless pumping in strongly interacting systems from a viewpoint of symmetry-protected topological phases. If the protecting symmetries of the underlying topological phases change between noninteracting fermions and a bosonic system in the strong coupling limit, the symmetry protection argument enforces a gap closing and thereby predicts a breakdown of the topological pumping. We also demonstrate that, even in the weakly interacting regime where the bulk topological pumping is still robust, the interaction effects manifest themselves in the edge density profiles, leading to a unique feature of the pumping in open boundary conditions. Furthermore, an extension of the above results indicates that an analog of an interaction-induced phase of Weyl semimetals can be realized in the setup of the topological pumping. 
Our results provide a systematic understanding for the stability of topological pumping against strong interactions, to which the conventional perturbative argument cannot be applied.
\end{abstract}


\maketitle


\section{Introduction}
Topological phases in strongly correlated systems have been of great interest in condensed matter physics over recent years. Although the understanding of topological phases was based on free-fermion systems at the primary stage \cite{TKNN,Kohmoto,KaneMele}, the notions of topological orders \cite{WenIntJMod,Wen17} and symmetry-protected topological (SPT) phases \cite{Chen13,SPTbook,Senthil15} have generalized the concepts to interacting many-body systems and have provided the theoretical framework. In parallel with the conceptual developments, numerous efforts have been devoted to uncover correlation effects on topological insulators in various systems
 \cite{Hohenadler13,WitczakKrempa13,Chiu16,Dzero16,Hohenadler11,Yamaji11,Yu11,Yoshida12,Alexandrov15,Peters16},
 and active debates are still continuing.

The topological nature of quantum states is not limited to the topological phases but also appears in time-dependent dynamics. The topological Thouless pumping \cite{Thouless83}, which is the main focus of this paper, is such a prototypical example. Here, an adiabatic cycle of a one-dimensional (1D) band insulator is considered. Then, the pumped charge per cycle is given by the Chern number, leading to quantized transport which arises from the topology of the ground state over the entire time evolution. Despite its significance, the topological pumping had not been observed in experiments until recently. However, it was finally realized using ultracold atoms in optical lattices \cite{Nakajima,Lohse16,Schweizer,Lohse18}. This remarkable progress has stimulated new theoretical proposals such as fractional pumping \cite{Zeng15,Zeng16,NakagawaFurukawa,Taddia17} and has further raised the importance of taking strong correlations into account, since the interactions of cold atoms are highly controllable.

The fundamental issue in the correlation effect on the free-fermion topological pumping is the stability of the pumping against interactions. In general, the stability is guaranteed for weak interactions since the (many-body) Chern number cannot be changed as long as the energy gap does not close during the cycle \cite{NiuThouless}, but systematic predictions for strong interactions are not available up to now. In this paper, we show that the underlying SPT phases in the pumping protocol shed light on this problem and provide a simple criterion for the stability of the topological pumping in the strongly interacting regime. The key observation for this is that an SPT phase changes from fermionic to bosonic character when increasing the interaction strength \cite{Manmana12,Yoshida14,YouXu14,Yoshida16,WuYoshida16,Bi17,NakagawaKawakami17}. Strong interactions freeze the motion of fermions at low energies, and the system is effectively described by a bosonic system as in the standard treatment of spin systems in Mott insulators. Reflecting this absence of the fermionic degrees of freedom, the protecting symmetries of the SPT phases may be different between the fermionic phase and the bosonic limit \cite{NakagawaKawakami17,AnfusoRosch,MoudgalyaPollmann}. We show that the emergent constriction of low-energy Hilbert space due to strong interactions and the change of the protecting symmetries 
enforce a gap closing during the cycle, thereby predicting the instability of the topological pumping. We also demonstrate that, even if weak interactions do not affect the bulk quantized pumping, the correlation effect is yet evident when imposing open boundary conditions. It causes the emergence of Mott insulating states at edges \cite{Yoshida14,Yoshida16} and offers a unique bulk-edge correspondence in topological pumping \cite{HatsugaiFukui16}. Finally, we point out that extending the pumping protocol by an additional adiabatic parameter gives an analog of an interaction-induced phase in Weyl semimetals \cite{MorimotoNagaosa}, by highlighting the fate of 
the Weyl point, the pumped charge, and the surface Fermi arcs 
in the interacting system. Our results are readily testable by the current experimental setup in ultracold atoms \cite{Nakajima,Wang13}.

The paper is organized as follows. In Sec.\ \ref{sec_model}, we introduce our model and explain the connection between the topological pumping and SPT phases focusing on the description of the pumping in terms of many-body polarization. Next, in Sec.\ \ref{sec_breakdown}, we show that the topological pumping breaks down in the strongly interacting regime using an analytical argument based on symmetry protection of topological phases. We confirm the analytical prediction of the breakdown by numerical calculations in Sec.\ \ref{sec_numerics}, and also point out a unique correlation effect on edge states in the topological pumping. In Sec.\ \ref{sec_bond}, we make one remark that the model has an intermediate regime with a long-range order, and we describe the pumping in that regime. In Sec.\ \ref{sec_attractive}, the case of attractive interactions is considered and we show the instability of the spin pumping. In Sec.\ \ref{sec_WeylMott}, we point out that the interacting Thouless pumping provides a realization of an interaction-induced phase of Weyl semimetals. Finally, we summarize our paper in Sec.\ \ref{sec_summary}. Two Appendixes are devoted to supplementary calculations on the stability of the topological pumping (Appendix \ref{sec_bosonization}) and on the symmetry protection of topological phases (Appendix \ref{sec_MPS}).

\section{Model\label{sec_model}}
\begin{figure}
\includegraphics[width=8.5cm]{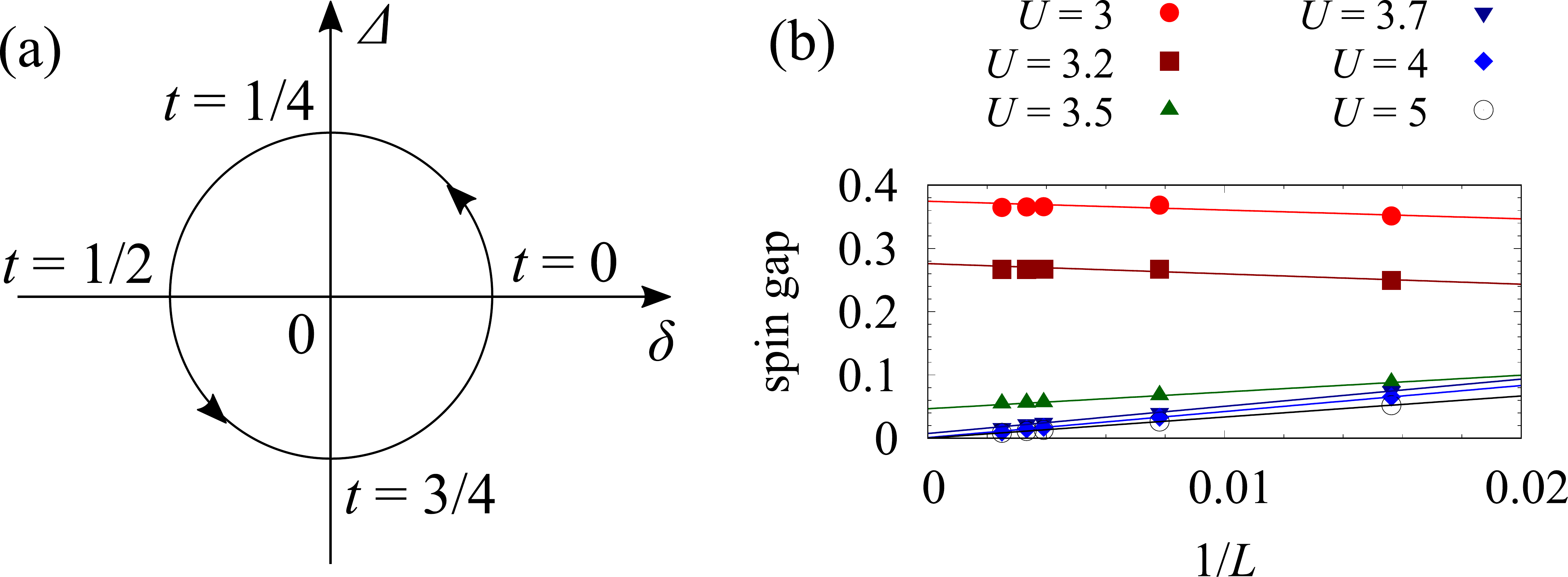}
\caption{(a) The pumping protocol of the Rice-Mele model. (b) Spin gap of the correlated Rice-Mele model at $\delta=0,\Delta=0.9$. The length of the chain is $L=64, 128, 256, 300,$ and $400$. Each line is a linear fit using three leftmost points.}
\label{fig_protocol}
\end{figure}

To elucidate the interaction effects on the Thouless pumping, we consider a spinful Rice-Mele model \cite{RiceMele} with the Hubbard interaction:
\begin{align}
H(t)=&-\sum_{j=0}^{L-1}\sum_{\sigma=\uparrow,\downarrow} (t_\mathrm{hop}+(-1)^j\delta(t))(c_{j\sigma}^\dag c_{j+1\sigma}+\mathrm{h.c.})\notag\\
&+\Delta(t)\sum_{j=0}^{L-1}\sum_{\sigma=\uparrow,\downarrow}(-1)^j c_{j\sigma}^\dag c_{j\sigma}
+U\sum_{j=0}^{L-1} n_{j\uparrow}n_{j\downarrow},
\label{eq_model}
\end{align}
where $U\geq0$. $c_{j\sigma}$ $(c_{j\sigma}^\dag)$ annihilates (creates) a fermion and $n_{j\sigma}=c_{j\sigma}^{\dag}c_{j\sigma}$ counts the particle number at site $j$. The ``time" $t$ is the adiabatic parameter and we set the time dependence as $\delta(t)=A\cos(2\pi t)$ and $\Delta(t)=A\sin(2\pi t)$ ($0\leq t \leq 1$) with $A>0$ (see Fig.\ \ref{fig_protocol} (a)). The on-site interaction $U$ is natural for ultracold atoms and the model \eqref{eq_model} can be realized by the setup in Refs.\ \onlinecite{Nakajima,Wang13}. Throughout the paper, we set $t_\mathrm{hop}=1$ as the unit of energy, and consider the model \eqref{eq_model} at half filling under the periodic boundary condition (PBC) $c_{L\sigma}=c_{0\sigma}$ or the open boundary condition (OBC) $c_{L\sigma}=0$. 

If the ground state stays gapped during the cycle, the pumped charge $Q$ over one cycle is given by the many-body Chern number\cite{NiuThouless}
\begin{align}
Q=\frac{1}{2\pi}\int_0^1dt\int_0^{2\pi}d\theta (&i\braket{\partial_t\Psi_\theta(t)|\partial_\theta\Psi_\theta(t)}\notag\\
&-i\braket{\partial_\theta\Psi_\theta(t)|\partial_t\Psi_\theta(t)}),
\label{eq_Chern}
\end{align}
where $\ket{\Psi_\theta(t)}$ is the many-body ground state of $H(t)$ under the twisted boundary condition with angle $\theta$. When the system is non-interacting, the integral over the angle $\theta$ can be replaced by the integral over the single-particle momentum of the Bloch states, and we obtain $Q=2$ for the model \eqref{eq_model} with $U=0$ by diagonalizing it after the Fourier transformation \cite{Xiao10}. The many-body Chern number, and thus the pumped charge, are rewritten by the change of the polarization \cite{NakagawaFurukawa,Watanabe18}. 
In the case of the PBC, the polarization $P_\sigma$ for spin $\sigma$ is defined by \cite{Resta98,RestaSorella99}
\begin{equation}
P_\sigma(t)=\frac{1}{2\pi}\mathrm{Im}\ln\bra{\Psi(t)}U_\sigma\ket{\Psi(t)} \ \ (\mathrm{mod}\ 1)
\label{eq_pol}
\end{equation}
with
\begin{gather}
U_\sigma=\exp\Bigl[\frac{2\pi i}{L}\sum_{j=0}^{L-1} (j-j_0)n_{j\sigma}\Bigr],\label{eq_twist}
\end{gather}
where $j_0=\frac{L-1}{2}$ is the center of the chain and $\ket{\Psi(t)}\equiv\ket{\Psi_{\theta=0}(t)}$ is the many-body ground state of $H(t)$ under the PBC. 
The exponentiation in Eq.\ \eqref{eq_twist} removes the ambiguity of the polarization associated with the PBC $j\sim j+L$ and the time derivative of the polarization leads to the current \cite{Watanabe18,Resta98}. 
The pumped charge $Q$ is then given by the change in the polarization 
\begin{gather}
Q=\int_0^1 dt (\partial_t P_\uparrow(t)+\partial_t P_\downarrow(t)).
\label{eq_pumpedcharge}
\end{gather}
Importantly, the polarization \eqref{eq_pol} is enforced to be quantized when some symmetries are imposed. For example, under the site-centered inversion symmetry by $I_sc_{j\sigma}I_s^\dag=c_{L-j\sigma}$, the polarization is quantized as $P_\sigma=1/4$ or $3/4$ since $I_sU_{\sigma}I_s^\dag=-U_\sigma^\dag$. Similarly, under the bond-centered inversion symmetry by $I_bc_{j\sigma}I_b^\dag=c_{L-j-1\sigma}$, the polarization is $P_\sigma=0$ or $1/2$ since $I_bU_{\sigma}I_b^\dag=U_\sigma^\dag$. In some cases, the quantized polarization can be used as a non-local ``order parameter" of SPT phases \cite{NakamuraTodo02,NakamuraVoit02}. In fact, since the model \eqref{eq_model} is reduced to the Su-Schrieffer-Heeger (SSH) model \cite{SSH} (up to the trivial spin degeneracy\footnote{In the original paper by Su, Schrieffer, and Heeger \cite{SSH}, electrons coupled with lattice distortion (phonons) were considered. Therefore, to be precise, what we call the SSH model corresponds to the model in Ref.\ \onlinecite{SSH} with replacing the phonon variable by its mean-field value.}) when $\Delta=0$ and $U=0$, the quantized polarization $P_\sigma=0$ corresponds to the trivial phase at $\delta>0$ and $P_\sigma=1/2$ corresponds to the SPT phase at $\delta<0$ protected by the bond-centered inversion symmetry\footnote{Our model \eqref{eq_model} with $\Delta=0$ has an accidental chiral (sublattice) symmetry. Although the SPT phase in the SSH model can also be protected by the chiral symmetry, we note that the chiral symmetry does not exist in real experimental situations, since the hopping between sites in the same sublattice is allowed. Therefore, the inversion symmetry is the most relevant symmetry which protects the SPT phase in generic situations.}. For obtaining a nonzero pumped charge, the inversion symmetry, which protects the quantized value of the polarization, must necessarily be broken during the pumping cycle. Thus the pumping protocol can be regarded as a process that smoothly connects the two distinct (SPT and trivial) phases in the SSH model without gap closing by symmetry-breaking perturbation $\Delta$. This picture of the topological pumping is widely applicable to various pumping schemes \cite{NakagawaFurukawa,Wang13,BergLevinAltman,Shindou}.

\section{Breakdown of the Thouless pumping\label{sec_breakdown}}
The topological pumping of the model \eqref{eq_model} is robust against small Hubbard $U$ since the quantized pumped charge cannot be changed as long as the energy gap does not close \cite{NiuThouless}. This can be also shown more explicitly using bosonization methods (see Appendix \ref{sec_bosonization}). However, we show that this topological pumping finally breaks down due to a gap closing by considering the strong coupling regime, where $|\delta|,|\Delta|,t_\mathrm{hop}\ll U$. In this regime, using a standard second-order perturbation theory (the Schrieffer-Wolff transformation), we obtain an effective spin chain from Eq.\ \eqref{eq_model} as
\begin{equation}
H_{\mathrm{eff}}=\sum_{j=0}^{L-1}(J+(-1)^j\delta')\bm{S}_j\cdot\bm{S}_{j+1},
\label{eq_spinPeierls}
\end{equation}
where $J=\frac{2t_\mathrm{hop}^2+2\delta^2}{U-2\Delta}+\frac{2t_\mathrm{hop}^2+2\delta^2}{U+2\Delta}$ and $\delta'=\frac{4t_\mathrm{hop}\delta}{U-2\Delta}+\frac{4t_\mathrm{hop}\delta}{U+2\Delta}$. Here $\bm{S}_j=\frac{1}{2}\sum_{\alpha\beta}c_{j\alpha}^\dag\bm{\sigma}_{\alpha\beta}c_{j\beta}$ is the spin operator. When $\delta'\neq0$, the ground state of the model \eqref{eq_spinPeierls} is the spin-Peierls phase with a finite spin gap \cite{Giamarchi}. In the SSH model (i.e. $\Delta=0$) with the Hubbard interaction, it is known that the ground state smoothly crossovers from the noninteracting fermionic phase to the spin-Peierls phase without closing the energy gap \cite{Manmana12,Yoshida14}. On the other hand, the model \eqref{eq_spinPeierls} at $\delta'=0$ is reduced to the spin-$1/2$ Heisenberg chain with a gapless ground state. The fermionic chain \eqref{eq_model} with $\delta=0$ is called the ionic Hubbard model \cite{Nagaosa86,Egami93}. In this case, the spin gap of the band insulator at $U=0$ is decreased with increasing Hubbard $U$, and finally it shows a phase transition into the gapless phase at a critical value $U=U_c\simeq 2|\Delta|$.
\cite{Nagaosa86,Egami93,RestaSorella95,Ortiz96,Fabrizio99,Torio01,Manmana04,OtsukaNakamura05} In Fig.\ \ref{fig_protocol} (b), we have calculated the spin gap of the model \eqref{eq_model} with $\delta=0, \Delta=0.9$ by the density-matrix renormalization group (DMRG) method \cite{White92,Schollwoeck05,Schollwoeck11} in the PBC keeping up to $800$ states. The spin gap $E(N,S_z=1)-E(N,S_z=0)$ is defined by the difference of ground-state energies in different spin quantum numbers $S_z$ with the fixed particle number $N=L/2$. Figure \ref{fig_protocol} (b) shows that the spin gap closes above $U\simeq 4$, being consistent with literature \cite{Torio01}. This fact implies that the topological pumping of the Rice-Mele model \eqref{eq_model} breaks down at $U>U_c$ due to the vanishing many-body gap at $t=1/4$ and $t=3/4$ during one pumping cycle.

We here show that this breakdown of topological pumping is not accidental in this specific model, but is a generic phenomenon caused by the underlying SPT phases in the pumping protocol. To see this, it is worth noting that the two distinct spin-Peierls phases at $\delta'>0$ and $\delta'<0$ are protected by either of the time-reversal, the bond-centered inversion, or the spin dihedral symmetries when we restrict the Hilbert space to the spin system \eqref{eq_spinPeierls}. This fact is proved by using the matrix-product-state formalism \cite{Chen1, Chen2} in Appendix \ref{sec_MPS}, but can be naturally understood by noticing that the ground states in extreme cases $\delta'=\pm J$ are a cousin of the valence-bond-solid state \cite{AKLT1,AKLT2}, which is the celebrated wavefunction of the Haldane phase \cite{Haldane1,Haldane2,GuWen}. The symmetry protection of the spin-Peierls phases then follows from that of the Haldane phase \cite{Pollmann10,Pollmann12}. Since the staggered potential $\Delta$ does not break the time-reversal and the spin-rotation symmetries, the symmetry protection of the spin-Peierls phases predicts that we cannot connect these distinct phases by $\Delta$ without gap closing or spontaneous breaking of all of the protecting symmetries. From this fact, the origin of the gap closing at $t=1/4$ and $t=3/4$ in the strongly correlated regime is understood as the symmetry protection of the spin-Peierls phase. Thus, we arrive at a conclusion that the breakdown of the Thouless pumping is caused by the change of the protecting symmetries of the SPT phase associated with the crossover from fermions to bosons. Since the (possible) case of spontaneous symmetry breaking also leads to a breakdown of pumping due to degenerate ground states, the Thouless pumping in the Rice-Mele model should break down at a certain critical $U$ if the protecting symmetries are not completely broken and the fermionic degrees of freedom become irrelevant at low energies in the strongly interacting regime.

\section{Numerical demonstration of the breakdown and correlation effects on edges\label{sec_numerics}}
Next, we demonstrate how the topological pumping breaks down by numerical calculations of the polarization using the DMRG method. 
Here we adopt the OBC and the polarization of the ground state $\ket{\Psi(t)}$ 
is defined by
\begin{equation}
P_{\mathrm{open}}(t)=\frac{1}{L}\sum_\sigma\sum_{j=0}^{L-1}\bra{\Psi(t)}(j-j_0)n_{j\sigma}\ket{\Psi(t)},
\label{eq_Popen}
\end{equation}
since there is no ambiguity of lattice coordinates like in the case of the PBC. This polarization is naturally equivalent to the center-of-mass position of particles and has been measured for detecting topological pumping in cold-atom experiments \cite{Nakajima,Lohse16}. 

\begin{figure}
\includegraphics[width=8.5cm]{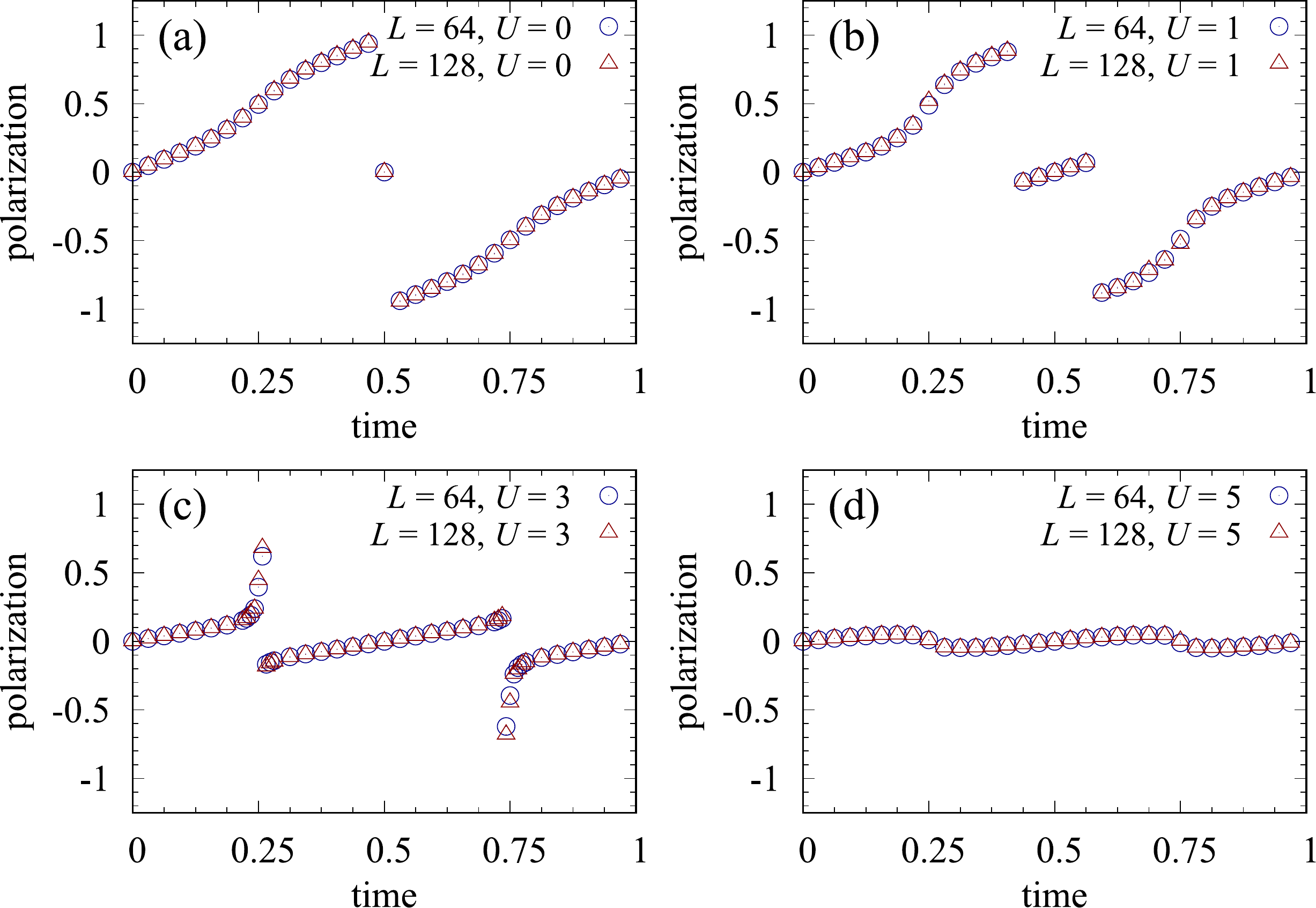}
\caption{Numerical results of the polarization $P_{\mathrm{open}}(t)$ of the correlated Rice-Mele model in the OBC. The parameter is $A=0.9$ and the length of the chain is $L=64,128$. The interaction strength is set as (a) $U=0$, (b) $U=1$, (c) $U=3$ and (d) $U=5$.}
\label{fig_polarization}
\end{figure}

We note that in the OBC, where hard walls exist at the boundary, the total change in the polarization is automatically zero. This fact is reconciled with the occurrence of the topological pumping through the following reasoning \cite{HatsugaiFukui16}. Let us first consider the non-interacting case. Then the model \eqref{eq_model} in the OBC possesses in-gap states localized around the edges. 
The emergence of the in-gap states is the consequence of the nontrivial Chern number of the bulk pumping protocol, where the non-interacting model \eqref{eq_model} can be mapped to a Chern insulator with chiral edge states when regarding the time as momentum for the second dimension. The edge states have zero energy at $t = 1/2$ and are four-fold degenerate at each edge in the SSH model \footnote{The zero-energy edge states at $t = 1/2$ can be regarded as the topologically protected edge modes in the SPT phase of the SSH model when the chiral symmetry is present. When only the inversion symmetry is present, the energy of the edge state in the SPT phase is not necessarily zero, since the inversion symmetry alone does not protect the zero-energy state \cite{Hughes11}. However, since the existence of the edge states is ensured by the nonzero Chern number, the absence of the chiral symmetry only leads to a shift of the time when the edge states have zero energy; the following discussion in the main text is not essentially changed by this effect.}. 
As a result, these edge states are occupied (unoccupied) in the left (right) edge at $t=1/2$, leading to a discontinuous change in polarization that compensates the bulk pumped charge as depicted in Fig.\ \ref{fig_polarization} (a). This mechanism is a manifestation of the bulk-edge correspondence in the topological pumping \cite{HatsugaiFukui16}. 

\begin{figure}
\includegraphics[width=8.5cm]{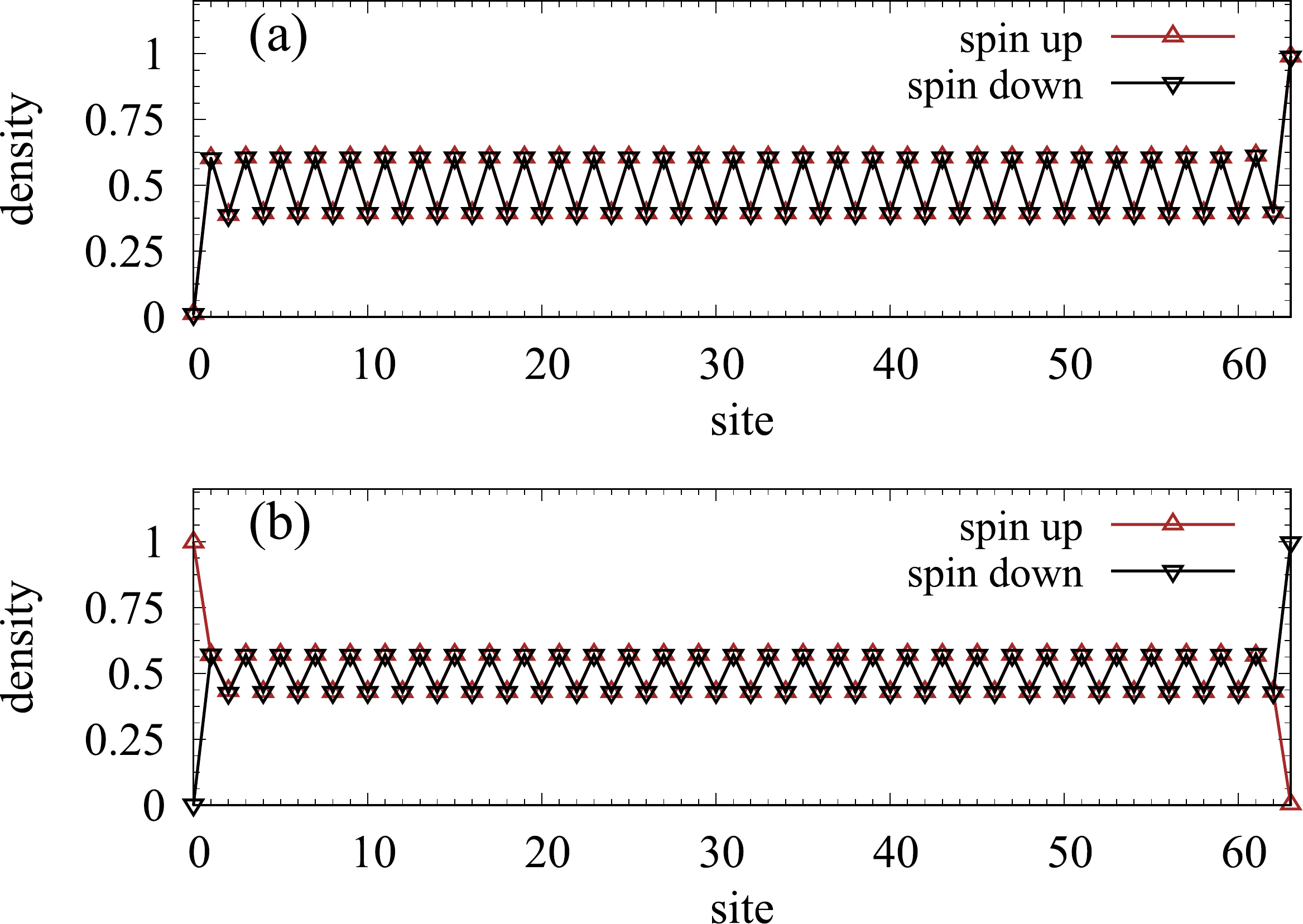}
\caption{Density profiles $\bra{\Psi(t)}n_{j\sigma}\ket{\Psi(t)}$ for each spin component. The parameters are $L=64, A=0.9, U=1$ and the times are taken as (a) $t=0.40625$ ($P_{\mathrm{open}}(t)=0.87909$) and (b) $t=0.43750$ ($P_{\mathrm{open}}(t)=-0.068911$). The discontinuous change in the polarization is caused by the change of occupation of the edge states.}
\label{fig_density}
\end{figure}

In the weakly interacting case with finite $U$ (Figs.\ \ref{fig_polarization}(b) and \ref{fig_polarization}(c)), we find that the bulk pumping is still robust while the discontinuous contribution from edge states splits into two pieces. The latter behavior is caused by the correlation effect in the edge state. In the vicinity of the SSH point, the low-lying excitation of the chain consists of the in-gap edge states, which are modeled by zero-dimensional Hubbard model  
\begin{equation}
H_\mathrm{edge}=\tilde{\Delta}\sum_{\sigma}(n_{L\sigma}-n_{R\sigma})+\tilde{U}(n_{L\uparrow}n_{L\downarrow}+n_{R\uparrow}n_{R\downarrow}).
\label{eq_0DHubbard}
\end{equation}
Here $L$ ($R$) denotes the left (right) edge, $\mathrm{sgn}(\tilde{\Delta})=\mathrm{sgn}(\Delta)$ and $\mathrm{sgn}(\tilde{U})=\mathrm{sgn}(U)$. 
When the Hubbard repulsion is switched on, the four-fold degenerate edge states of the SSH model immediately acquire a charge gap and two-fold degenerate edge spin states are left at each edge \cite{Yoshida14}. The emergent Mott insulating state at the edge persists at small $|\Delta|$, and correspondingly the changes of occupation at the edge occur at $2|\Delta|\simeq U$ where the edge charge gap vanishes. We verify this picture from the density profiles around the polarization discontinuity (Figs.\ \ref{fig_density}(a) and \ref{fig_density}(b)). 
The two-step reconstruction of the edge-state occupation offers a unique bulk-edge correspondence in the topological pumping which manifests the correlation effect even in the weakly interacting regime.

In the breakdown regime $U>U_c$ where the spin gap is closed, we confirm that the polarization does not exhibit a discontinuous change and the charge pumping does not occur (Fig.\ \ref{fig_polarization} (d)). 
We note that this result is not trivial since the analytical argument in the previous section only states that the pumped charge becomes non-quantized due to the vanishing many-body gap above the critical interaction strength. Nevertheless, the numerical result in Fig.\ \ref{fig_polarization} (d) shows that not only the gapped pumping process breaks down, but also the pumped charge totally vanishes although it can slightly deviate from zero due to the breakdown of adiabaticity. 
The small amplitude of the change in the polarization is reasonable since in the deep Mott insulating limit $U\to\infty$ the charge density is frozen so that $\sum_\sigma n_{j\sigma}=1$. 

\section{Pumping in the regime of bond order\label{sec_bond}}

\begin{figure*}[t]
\includegraphics[width=16cm]{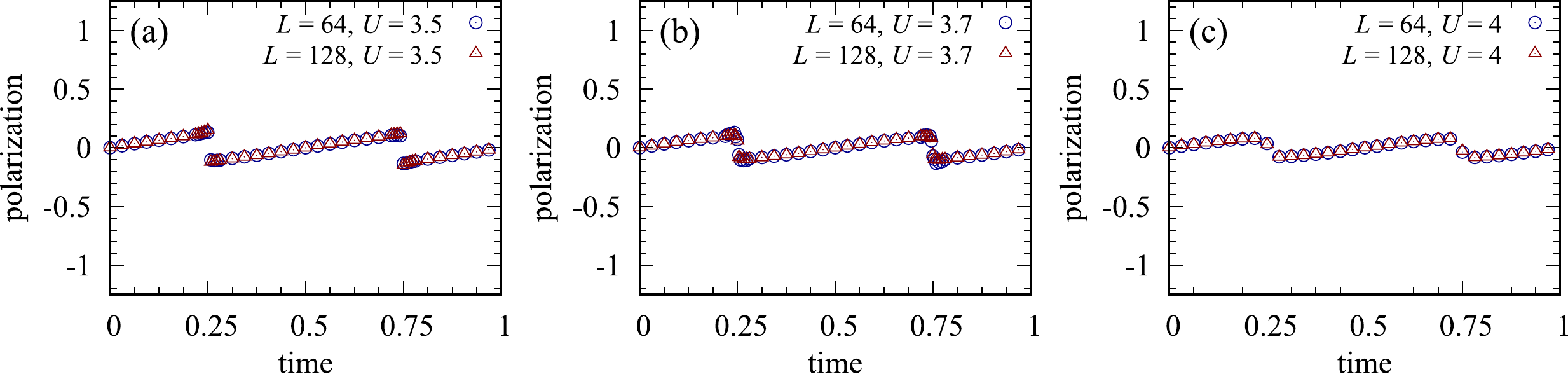}
\caption{Numerical results of the polarization $P_{\mathrm{open}}(t)$ of the correlated Rice-Mele model in the OBC. The parameter is $A=0.9$ and the length of the chain is $L=64, 128$. The interaction strength is set as (a) $U=3.5$, (b) $U=3.7$, and (c) $U=4$.}
\label{fig_BO}
\end{figure*}

We make here one important comment that, in the ionic Hubbard model, the direct transition from the gapped phase to the gapless phase is intervened by a bond-ordered phase which emerges at $U_b<U<U_c$ by spontaneously breaking the site-centered inversion symmetry \cite{Fabrizio99,Torio01,Manmana04,OtsukaNakamura05}. 
The bond-order parameter for the ground state $\ket{\Psi}$ is given by
\begin{align}
\langle B\rangle &\equiv \frac{1}{L}\sum_{\sigma}\sum_{j=0}^{L-1}(-1)^j\bra{\Psi} c_{j\sigma}^\dag c_{j+1\sigma}+c_{j+1\sigma}^\dag c_{j\sigma}\ket{\Psi}\notag\\
&=\frac{1}{L}\bra{\Psi}\frac{\partial H}{\partial \delta}\ket{\Psi},
\end{align}
and 
\begin{equation}
\langle B\rangle|_{\delta\to +0}= -\langle B\rangle|_{\delta\to -0}\neq 0
\label{eq_BOjump}
\end{equation}
in the bond-ordered phase. In Figs.\ \ref{fig_BO}(a) and \ref{fig_BO}(b), we plot the evolution of the polarization in the correlated Rice-Mele model in the regime where the bond order appears with $\delta=0$. The discontinuous jumps at $t=1/4$ and $t=3/4$ signal the bond order due to Eq.\ \eqref{eq_BOjump}. Since the ground states in the bond-ordered phase are doubly degenerate and break the closed pumping process, the pumped charge in these cases is not quantized. Associated with the decrease of the bond order parameter by approaching $U=U_c$, the non-quantized pumping smoothly changes into the breakdown regime (see Figs.\ \ref{fig_BO}(b) and \ref{fig_BO}(c)). 

Although this breakdown mechanism is not related to the symmetry protection and thus may not be generic, we expect that this phenomenon is relevant for experiments combining a scheme to probe the bond order using superlattice modulation spectroscopy \cite{Loida17}.

\section{Attractive interactions, spin pumping\label{sec_attractive}}
\begin{figure}[b]
\includegraphics[width=8.5cm]{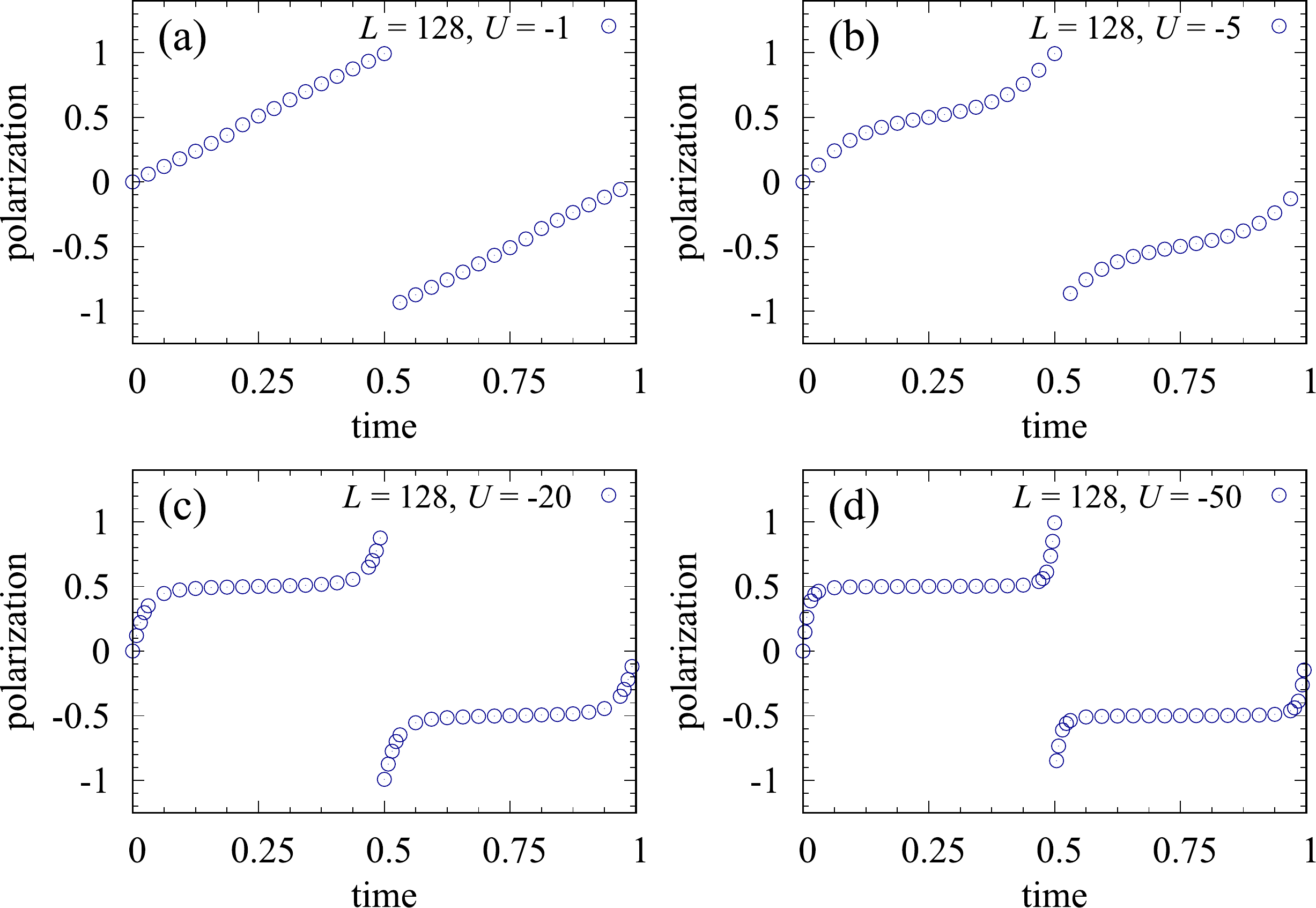}
\caption{Numerical results of the polarization $P_{\mathrm{open}}(t)$ of the Rice-Mele model in the OBC with attractive interactions. The parameter is $A=0.9$ and the length of the chain is $L=128$. The interaction strength is set as (a) $U=-1$, (b) $U=-5$, (c) $U=-20$ and (d) $U=-50$.}
\label{fig_attractive}
\end{figure}

Our stability argument can be applied to other types of interactions and pumping protocols. For example, let us consider the attractive interaction $U<0$ in the model \eqref{eq_model}. 
In Fig.\ \ref{fig_attractive}, we plot the DMRG results of the polarization in the case of the attractive interactions. We numerically confirm up to $U=-50$ that the bulk quantized pumping persists under the attractive interaction. 
This result indicates that the topological charge pumping is robust against any $U<0$. This is reasonable since the SSH model of fermions approaches that of hard-core bosons in the $U\to-\infty$ limit and the protecting symmetry does not change \footnote{To be precise, the SPT phase in the SSH model of hard-core bosons is also protected by a charge-rotation symmetry which is connected to the spin-rotation symmetry via the Shiba transformation, but this symmetry is broken by the staggered potential.} 
We note that the discontinuous jump of polarization due to the edge contribution always takes place at $t=1/2$ in these data. This behavior is consistent with the effective description of the edge states by the zero-dimensional Hubbard model \eqref{eq_0DHubbard}, 
since the ground state of $H_\mathrm{edge}$ in the half-filling sector is the $\ket{n_L=0,n_R=2}$ state for $\tilde{\Delta}>0$ and the $\ket{n_L=2,n_R=0}$ state for $\tilde{\Delta}<0$ (here $n_\alpha\equiv n_{\alpha\uparrow}+n_{\alpha\downarrow}$).

On the other hand, performing the Shiba transformation\cite{Shiba} $c_{j\uparrow}^{\dag}\to c_{j\uparrow}^\dag,c_{j\downarrow}^{\dag}\to e^{i\pi j}c_{j\downarrow}$, we find that the model \eqref{eq_model} at half-filling is equivalent to
\begin{align}
H&(t)=-\sum_{j=0}^{L-1}\sum_{\sigma=\uparrow,\downarrow} (t_\mathrm{hop}+(-1)^j\delta(t))(c_{j\sigma}^\dag c_{j+1\sigma}+\mathrm{h.c.})\notag\\
&+\Delta(t)\sum_{j=0}^{L-1} (-1)^j (c_{j\uparrow}^\dag c_{j\uparrow}-c_{j\downarrow}^\dag c_{j\downarrow})
-U\sum_{j=0}^{L-1} n_{j\uparrow}n_{j\downarrow},
\label{eq_spinpump}
\end{align}
and this model describes the spin pumping \cite{Shindou,FuKane} with the inverted sign of the interaction. Therefore, the spin pumping is fragile to strong attractive interactions while it is robust against repulsive interactions. Since the charge gap and the spin gap are interchanged by the Shiba transformation, the breakdown of the spin pumping is caused by the gap closing in the charge sector. The gapless phase appearing in $U>U_c$ with $\delta=0$ corresponds to a Luttinger liquid of molecular bosons formed by the fermion pairs. The robustness of the spin pumping against the repulsive interaction is consistent with the symmetry protection of the spin-Peierls phase, since the staggered magnetic field $\Delta$ in Eq.\ \eqref{eq_spinpump} breaks all of the protecting symmetries.

\begin{figure}
\includegraphics[width=8.5cm]{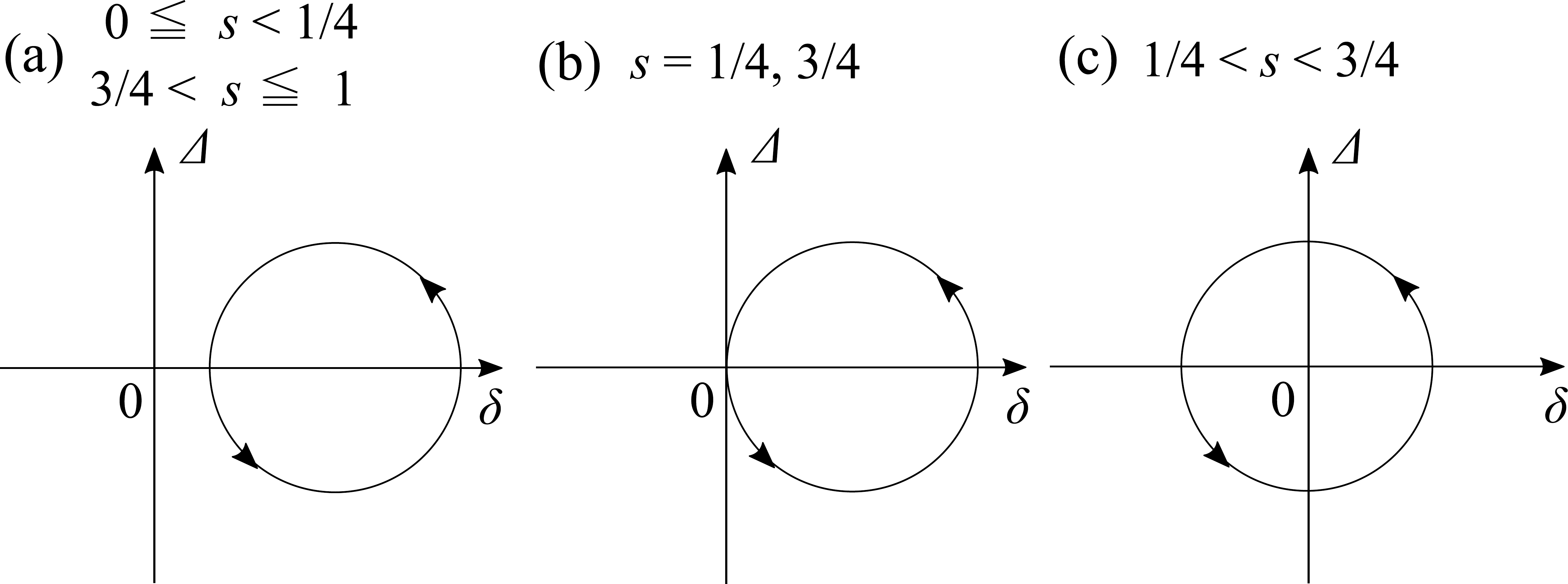}
\caption{Pumping protocol for the topological-pumping realization of a Weyl semimetal. The circles represent the trajectories of the parameters when $t$ is varied with $s$ being fixed. Each figure corresponds to the case of (a) $0\leq s<1/4, 3/4<s\leq1$, (b) $s=1/4,3/4$, and (c) $1/4<s<3/4$.}
\label{fig_protocolWeylMott}
\end{figure}

\section{Topological-pumping realization of Weyl Mott insulators\label{sec_WeylMott}}
Finally, as a byproduct of our results on the correlated topological pumping, we show that an extension of the present setup provides an analog of an interaction-induced phase of Weyl semimetals\cite{MorimotoNagaosa}. 
A Weyl semimetal is constructed by stacking two-dimensional band insulators in the momentum space. When the Chern number $C(k_z)$ parametrized in the stacking direction $k_z$ changes at a specific point, a Weyl fermion appears as a gap closing point in the band dispersion \cite{Wan11}. 
Since the gap closing is protected by the change of the Chern number, the Weyl fermion is topologically stable as long as non-interacting systems are considered, while the stability against interactions is not yet fully understood. In this context, Ref.\ \onlinecite{MorimotoNagaosa} proposed an intriguing possibility that an interaction opens a Mott gap at the Weyl points without spoiling the topological properties and drives the Weyl semimetal into a new phase called ``Weyl Mott insulator (WMI)". It is a correlation-driven insulator with gapped charge excitations but still hosts anomalous properties originating from the topological nature of the Weyl points. A WMI is characterized by (i) a charge gap in the single-particle excitation spectrum, (ii) nonvanishing Hall conductivity linked to the Chern number, and (iii) the surface Fermi arc which connects the Weyl points in the momentum space. While it was shown in Ref.\ \onlinecite{MorimotoNagaosa} that a momentum-space-decoupled interaction $U\sum_{\bm{k}}n_{\bm{k}\uparrow}n_{\bm{k}\downarrow}$  opens the Mott gap at the Weyl point for arbitrary $U>0$ and realizes the WMI, its existence has not been confirmed in real materials nor in model calculations with realistic interactions. 

\begin{figure}
\includegraphics[width=8.5cm]{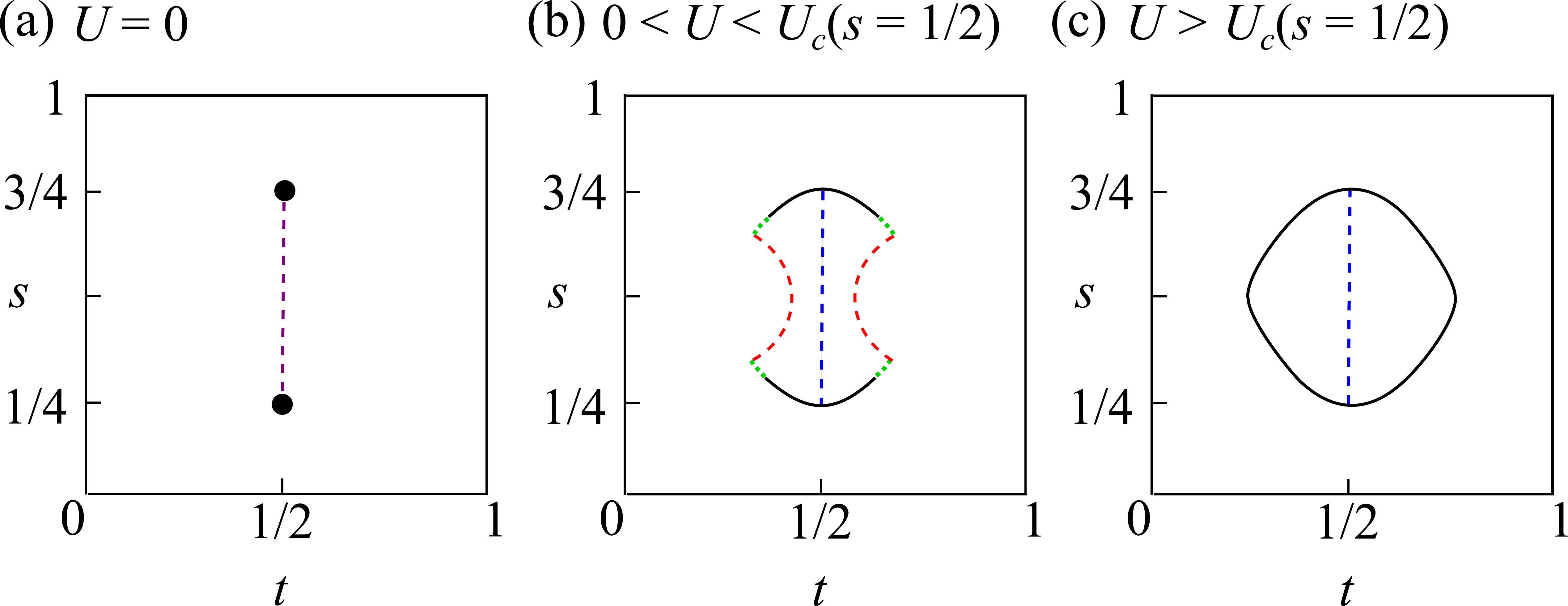}
\caption{Qualitative features of gapless lines in the $(t,s)$ space under the OBC for the 1D real space. The filled circles denote the Weyl points, and the solid lines indicate the parameters at which the bulk spin excitations are gapless. The broken red (blue) lines show the parameters at which the edge charge (spin) excitations are gapless. In (a), the  charge and spin excitations at the edge are degenerate, drawn by the broken purple line. The green dotted lines in (b) show the region where the bond order takes place. The interaction strengths are (a) $U=0$, (b) $0<U<U_c(s=1/2)$, and (c) $U>U_c(s=1/2)$.}
\label{fig_Fermiarc}
\end{figure}

A situation analogous to the Weyl semimetal is realized in the Thouless pumping by introducing an additional adiabatic parameter $s$ in the Hamiltonian, so that the Chern number changes at a specific value of $s$ (see also Refs.\ \onlinecite{DWZhang15, YBYang18}). 
For example, by setting parameters in Eq.\ \eqref{eq_model} as $\delta(t,s)=A[\cos(2\pi t)+\cos(2\pi s)+1]$ and $\Delta(t,s)=A\sin(2\pi t)$, the Chern number is $C(s)=2$ for $1/4<s<3/4$, and $C(s)=0$ for $0\leq s<1/4$ and  $3/4<s\leq 1$ (here $0\leq s\leq 1$) when $U=0$. 
See Fig.\ \ref{fig_protocolWeylMott}. 
Hence in the 3D ``Brillouin zone" spanned by $(k_x,t,s)$, where $k_x$ is the crystal momentum for the 1D real space, two gapless points emerge at $(\pi,1/2,1/4)$ and $(\pi,1/2,3/4)$ corresponding to the Weyl fermions (see Fig.\ \ref{fig_Fermiarc} (a)). 

Let us discuss the stability of the Weyl points against the Hubbard interaction in the model \eqref{eq_model}. 
Since $\delta(t,s)=\Delta(t,s)=0$ at these gapless points, the Weyl points are gapped out by the interaction for arbitrary $U>0$ and a gapless spin collective mode is left over, as is well-known in the 1D Hubbard physics \cite{Giamarchi}. 
On the other hand, since the topological pumping is stable against small $U$, the pumped charge integrated over $t$ and $s$ (the anomalous Hall conductivity in the Weyl semimetal) remains finite. 
This implies that the model for small $U>0$ realizes the WMI in the setup of the topological pumping. 

As shown in Secs.\ \ref{sec_breakdown} and \ref{sec_numerics}, the topological pumping for given $s$ breaks down above the critical interaction strength $U_c(s)$ due to closing of the spin gap at two specific times during the cycle. Since $U_c(s)$ vanishes when the cycle crosses the Weyl point (i.e. $s=1/4,3/4$), the topological pumping near the Weyl point breaks down for $U>0$ and the associated gapless points form a line starting from the Weyl point in the $(t,s)$ plane, as shown in Fig.\ \ref{fig_Fermiarc} (b). Note that the gapless points represent collective spin excitations and the single-particle charge gap is finite in the whole $(t,s)$ plane in Fig.\ \ref{fig_Fermiarc} (b). In Fig.\ \ref{fig_Fermiarc}, we also show the parameters at which the edge states under the OBC for the 1D real space become gapless. The broken line for the gapless edge states at $U=0$ corresponds to the surface Fermi arc, on which the charge and spin modes are degenerate (see Fig.\ \ref{fig_Fermiarc} (a)). When the interaction is turned on, the surface Fermi arc connecting the Weyl points acquires a charge gap and only gapless spin excitations remain, since the edge states consist of those of the SSH model (see Sec.\ \ref{sec_numerics}). Furthermore, the two-step charge-gap closing which was demonstrated in Sec.\ \ref{sec_numerics} offers two new arcs of gapless charge excitations shown by the broken red lines in Fig.\ \ref{fig_Fermiarc} (b). The two new arcs merge into the bulk spectrum at the bond-order transition point (see Sec.\ \ref{sec_bond}), where the neutral charge gap closes and the single-particle gap remains finite\cite{Manmana04}. Such spin-charge separation of the surface Fermi arcs is a new interaction-enabled feature of the WMI in this model, which is absent in the original proposal\cite{MorimotoNagaosa}. When the interaction strength is increased, the breakdown region gradually expands and the integrated pumped charge decreases. Finally, the strong interaction destroys the charge pumping in the whole region of $1/4<s<3/4$, and the system turns into a trivial Mott insulator with vanishing anomalous Hall conductivity (see Fig.\ \ref{fig_Fermiarc} (c)). 

For comparison with the original proposal in Ref.\ \onlinecite{MorimotoNagaosa}, we note that the interaction in our model is independent of the adiabatic parameters $t,s$ but local in the 1D real space. We also remark that, in contrast to Ref.\ \onlinecite{MorimotoNagaosa}, the Chern number in our case is defined \textit{not} in the whole parameter space, since it is not well-defined in the bond-ordered regime and the breakdown regime. Although the possibility of the WMI with local interactions in three dimensions still remains elusive, the realization with topological pumping here gives a new example of the WMI other than the original proposal\cite{MorimotoNagaosa}.

\section{Discussion and summary\label{sec_summary}}
In summary, we have shown that the topological pumping generally breaks down in the strongly interacting regime if the protecting symmetries change between the fermionic and bosonic SPT phases which underlie the pumping protocol. 
In general, the stability of topological pumping against interactions depends on specific pumping schemes. For example, let us consider a simple ``sliding-lattice" scheme where particles in a periodic potential in the continuous space are considered and the protocol consists of shifting the potential with a constant velocity\cite{Thouless83, Lohse16}. In this case, since the time evolution over one cycle is equivalent to a spatial-translation operation of the entire system, the charge pumping should occur regardless of the interaction strength. Similarly, in charge pumping induced by flux threading in two-dimensional topological phases \cite{NakagawaFurukawa,Laughlin81,QiZhang08,Ran08,Assaad13}, the many-body gap can be expected to not collapse during the pumping cycle since the flux cannot change the bulk properties. As for the topological pumping in the Rice-Mele model (1), we have shown that the charge pumping breaks down for strong repulsive interaction due to the gap closing in the cycle, whereas it is stable against attractive interaction. What distinguishes the repulsively interacting Rice-Mele model from the other stable pumping schemes? In this paper, we have unveiled that the underlying 1D SPT phases in the pumping protocol and the change of their nature from fermions to bosons are the key to understanding the stability of the pumping against interactions. 
We note that the existence of a Mott phase in the bulk is not sufficient for the breakdown of the pumping, since the charge pumping is robust even in the deep Mott regime in the above sliding-lattice scheme and in a Bose-Mott insulator in the Rice-Mele model \cite{Lohse16} (the hard-core limit of which corresponds to the $U\to -\infty$ limit of our model). Since our argument for the mechanism of the breakdown is based on the symmetry, the criterion for the stability is not limited to the specific model considered in this paper but is also applicable to various models.

Furthermore, we have elucidated that the correlation effect on the topological pumping is not limited to the breakdown of the pumping. For weak repulsive interactions, while the bulk quantized pumping is still robust, the correlation effect opens a charge gap in edge states of the SSH model \cite{Yoshida14}. The emergence of the Mott insulating state at the edge induces a two-step reconstruction of the occupation of the edge states to be consistent with the bulk charge pumping. This behavior manifests a bulk-edge correspondence in the correlated topological pumping, extending the result of the previous work \cite{HatsugaiFukui16} to the interacting system.

Our results are directly relevant for analyzing the interaction effect on the cold-atom realization of topological pumping, and furthermore are expected to serve for a systematic understanding of the stability of various pumping schemes. 
Although we have focused on the adiabatic limit in this paper, extension of our results by considering non-adiabatic effects\cite{Rossini13, Lychkovskiy17} in conjunction with Floquet theory\cite{Shih94,Lindner17,Privitera18} will be an interesting subject for future investigation.

\begin{acknowledgments}
We are grateful to Shuta Nakajima, Azusa Sawada, Masaki Tezuka, and Yoshiro Takahashi for valuable discussions.
This work was supported by JSPS KAKENHI (Grants No.\ JP16K05501, No.\ JP18H01140, No.\ JP18H04316, No.\ JP18K03511, and No.\ JP18H05842) and a Grand-in-Aid for Scientific Research on Innovative Areas ``Topological Materials Science" (Grant No.\ JP15H05855). M.N. was supported by RIKEN Special Postdoctoral Researcher Program.
The numerical calculations were performed on supercomputer at the ISSP in the University of Tokyo.
\end{acknowledgments}

\appendix

\section{Bosonization analysis of the stability of the Thouless pumping against weak interactions\label{sec_bosonization}}
The stability of the topological pumping of the model (1) against weak interactions is explicitly shown by bosonization \cite{Giamarchi} valid in the weak-coupling regime $|\delta|, |\Delta|, U\ll t_\mathrm{hop}$. The low-energy effective theory is given by
\begin{align}
H=&\sum_{\xi=c,s}\frac{1}{2\pi}\int dx(v_\xi K_\xi(\nabla\theta_\xi(x))^2+\frac{v_\xi}{K_\xi}(\nabla\phi_\xi(x))^2)\notag\\
&+\frac{g}{\pi\alpha}\int dx\Bigl[\cos(\sqrt{2}\phi_c(x)-\gamma)\cos\sqrt{2}\phi_s(x)\Bigr]\notag\\
&+\frac{U}{2\pi^2\alpha}\int dx\Bigl[\cos 2\sqrt{2}\phi_c(x)-\cos 2\sqrt{2}\phi_s(x)\Bigr],
\label{eq_Hbos}
\end{align}
where $c_{j\sigma}=\frac{1}{\sqrt{2\pi\alpha}}(e^{i\pi j/2}e^{i(\theta_\sigma(x)-\phi_\sigma(x))}+e^{-i\pi j/2}e^{i(\theta_\sigma(x)+\phi_\sigma(x))}), g=\sqrt{4\delta^2+\Delta^2}$ and $\gamma=\arctan\frac{\Delta}{2\delta}$.
The charge (spin) mode is defined by $\phi_c\equiv\frac{1}{\sqrt{2}}(\phi_\uparrow+\phi_\downarrow)$ $(\phi_s\equiv\frac{1}{\sqrt{2}}(\phi_\uparrow-\phi_\downarrow))$. Since the Umklapp $U$ term in Eq.\ \eqref{eq_Hbos} is less relevant than the $g$ term in the renormalization-group sense, the low-energy behavior is mainly governed by the second line in Eq.\ \eqref{eq_Hbos}, thereby pinning the boson fields at the potential minimum. We note that the pinning position of the charge boson field is interpreted as the charge polarization of the ground state \cite{Shindou, NakamuraVoit02}. When the adiabatic cycle is performed, the pinning position of the charge mode $\phi_c$ changes by $\pi\sqrt{2}$ and this change induces the quantized charge pumping $Q=\frac{\sqrt{2}}{\pi}\int_0^1 dt\partial_t\phi_c=2$.\cite{Shindou, BergLevinAltman} This analysis indicates that the Thouless pumping is stable against the Hubbard interaction at least in the weak $U$ region. We emphasize that this analysis is applicable only to the weak-coupling regime. In the strong coupling regime $|\delta|,|\Delta|,t_\mathrm{hop}\ll U$, the charge mode is gapped out and the effective theory is composed of only the spin mode. In this case, the charge pumping cannot occur. In this bosonization picture, the breakdown of the topological pumping can be understood as a consequence of the competition between the second line and the third line in Eq.\ \eqref{eq_Hbos}, which cannot be simultaneously minimized if $\gamma\neq 0,\pi$.\cite{NakagawaKawakami17}

\section{Symmetry protection of the spin-Peierls phase\label{sec_MPS}}
The symmetry protection of the spin-Peierls phase is proved by using the matrix-product-state (MPS) representation of the ground state \cite{Chen1, Chen2}. The unit cell of the spin chain (5) consists of two sites. The Hilbert space for the unit cell is therefore spanned by a basis set $\{ \ket{\uparrow\uparrow},\ket{\uparrow\downarrow},\ket{\downarrow\uparrow},\ket{\downarrow\downarrow}\}$. Since the local singlet pair is expressed as
\begin{gather}
\ket{\uparrow\downarrow}-\ket{\downarrow\uparrow}=\sum_{i_1,i_2=\uparrow,\downarrow}\mathrm{Tr}[A_{i_1}^{[1]}A_{i_2}^{[2]}]\ket{i_1i_2},
\end{gather}
with
\begin{align}
A_{\uparrow}^{[1]}&=
\begin{pmatrix}
0 & 1\\
0 & 0
\end{pmatrix},\ 
A_{\downarrow}^{[1]}=
\begin{pmatrix}
-1 & 0\\
0 & 0
\end{pmatrix},\\
A_{\uparrow}^{[2]}&=
\begin{pmatrix}
1 & 0\\
0 & 0
\end{pmatrix},\ 
A_{\downarrow}^{[2]}=
\begin{pmatrix}
0 & 0\\
1 & 0
\end{pmatrix},
\end{align}
the spin-Peierls ground states in the extreme cases $\delta'=\pm J$ are written as
\begin{gather}
\ket{\Psi}=\sum_{j_1,\cdots,j_{L/2}=\uparrow\uparrow,\uparrow\downarrow,\downarrow\uparrow,\downarrow\downarrow}\mathrm{Tr}[B_{j_1}\cdots B_{j_{L/2}}]\ket{j_1j_2\cdots j_{L/2}},
\end{gather}
where
\begin{align}
B_{\uparrow\uparrow}&=A_{\uparrow}^{[1]}A_{\uparrow}^{[2]}=0,\label{eq_Buu1}\\
B_{\uparrow\downarrow}&=A_{\uparrow}^{[1]}A_{\downarrow}^{[2]}=
\begin{pmatrix}
1 & 0\\
0 & 0
\end{pmatrix},\\
B_{\downarrow\uparrow}&=A_{\downarrow}^{[1]}A_{\uparrow}^{[2]}=
\begin{pmatrix}
-1 & 0\\
0 & 0
\end{pmatrix},\\
B_{\downarrow\downarrow}&=A_{\downarrow}^{[1]}A_{\downarrow}^{[2]}=0,
\end{align}
for $\delta'=J$, and
\begin{align}
B_{\uparrow\uparrow}&=A_{\uparrow}^{[2]}A_{\uparrow}^{[1]}=
\begin{pmatrix}
0 & 1\\
0 & 0
\end{pmatrix},\\
B_{\uparrow\downarrow}&=A_{\uparrow}^{[2]}A_{\downarrow}^{[1]}=
\begin{pmatrix}
-1 & 0\\
0 & 0
\end{pmatrix},\\
B_{\downarrow\uparrow}&=A_{\downarrow}^{[2]}A_{\uparrow}^{[1]}=
\begin{pmatrix}
0 & 0\\
0 & 1
\end{pmatrix},\\
B_{\downarrow\downarrow}&=A_{\downarrow}^{[2]}A_{\downarrow}^{[1]}=
\begin{pmatrix}
0 & 0\\
-1 & 0
\end{pmatrix},\label{eq_Bdd2}
\end{align}
for $\delta'=-J$. The MPS transforms as
\begin{align}
B_j&\stackrel{T}{\to}e^{i\vartheta_T}U_T^\dag B_j U_T,\\
B_j&\stackrel{I}{\to}e^{i\vartheta_I}U_I^\dag B_j U_I,\\
B_j&\stackrel{R_x}{\to}e^{i\vartheta_x}U_x^\dag B_j U_x,\\
B_j&\stackrel{R_z}{\to}e^{i\vartheta_z}U_z^\dag B_j U_z,
\end{align}
under the time-reversal operation $T$, the bond-centered inversion $I$, and the spin $\pi$ rotation $R_x$ ($R_z$) around $x$ ($z$) axis, respectively. For the spin-Peierls states, Eqs.\ \eqref{eq_Buu1}-\eqref{eq_Bdd2} lead to
\begin{align}
U_T=U_I=U_x=U_z=1
\end{align}
for $\delta'=J$, and
\begin{align}
U_T=U_I=
\begin{pmatrix}
0 & -i\\
i & 0
\end{pmatrix},\ 
U_x=
\begin{pmatrix}
0 & 1\\
1 & 0
\end{pmatrix},\ 
U_z=
\begin{pmatrix}
1 & 0\\
0 & -1
\end{pmatrix}
\end{align}
for $\delta'=-J$. The SPT invariant for each symmetry is given by \cite{Pollmann10, Pollmann12}
\begin{align}
\mathcal{O}_T&=\mathrm{Tr}[U_TU_T^*]/\chi,\\
\mathcal{O}_I&=\mathrm{Tr}[U_IU_I^*]/\chi,\\
\mathcal{O}_{xz}&=\mathrm{Tr}[U_xU_zU_x^\dag U_z^\dag]/\chi,
\end{align}
($\chi$ is the dimension of the matrices) and we obtain $\mathcal{O}_T=\mathcal{O}_I=\mathcal{O}_{xz}=1$ for $\delta'=J$ and $\mathcal{O}_T=\mathcal{O}_I=\mathcal{O}_{xz}=-1$ for $\delta'=-J$. This completes the proof of the symmetry protection of the two distinct spin-Peierls phases.

\bibliography{Corr_pump_ref.bib}

\end{document}